\documentclass[aps,twocolumn,amsmath,amssymb,showpacs,prb,superscriptaddress,unsortedaddress]{revtex4}
\usepackage{epsf}
\usepackage[pdftex]{graphicx}

\newcommand{\etal}{{\it et al.}}

\begin{document}

\title{Surface-driven electronic structure in LaFeAsO studied by angle resolved photoemission spectroscopy}

\author{Chang~Liu}
\affiliation{Division of Materials Science and Engineering, Ames Laboratory, Ames, Iowa 50011, USA}
\affiliation{Department of Physics and Astronomy, Iowa State University, Ames, Iowa 50011, USA}

\author{Yongbin~Lee}
\affiliation{Division of Materials Science and Engineering, Ames Laboratory, Ames, Iowa 50011, USA}

\author{A.~D.~Palczewski}
\affiliation{Division of Materials Science and Engineering, Ames Laboratory, Ames, Iowa 50011, USA}
\affiliation{Department of Physics and Astronomy, Iowa State University, Ames, Iowa 50011, USA}

\author{J.~-Q.~Yan}
\affiliation{Division of Materials Science and Engineering, Ames Laboratory, Ames, Iowa 50011, USA}

\author{Takeshi~Kondo}
\affiliation{Division of Materials Science and Engineering, Ames Laboratory, Ames, Iowa 50011, USA}
\affiliation{Department of Physics and Astronomy, Iowa State University, Ames, Iowa 50011, USA}

\author{B.~N.~Harmon}
\affiliation{Division of Materials Science and Engineering, Ames Laboratory, Ames, Iowa 50011, USA}
\affiliation{Department of Physics and Astronomy, Iowa State University, Ames, Iowa 50011, USA}

\author{R.~W.~McCallum}
\affiliation{Division of Materials Science and Engineering, Ames
Laboratory, Ames, Iowa 50011, USA} \affiliation{Department of
Materials Science and Engineering, Iowa State University, Ames, Iowa
50011, USA}

\author{T.~A.~Lograsso}
\affiliation{Division of Materials Science and Engineering, Ames Laboratory, Ames, Iowa 50011, USA}

\author{A.~Kaminski}
\affiliation{Division of Materials Science and Engineering, Ames Laboratory, Ames, Iowa 50011, USA}
\affiliation{Department of Physics and Astronomy, Iowa State University, Ames, Iowa 50011, USA}

\date{\today}
\begin{abstract}

We measured the electronic structure of an iron arsenic parent
compound LaFeAsO using angle resolved photoemission spectroscopy
(ARPES). By comparing with a full-potential Linear Augmented Plane
Wave calculation we show that the extra large $\Gamma$ hole pocket
measured via ARPES comes from electronic structure at the sample
surface. Based on this we discuss the strong polarization dependence
of the band structure and a temperature-dependent hole-like band
around the $M$ point. The two phenomena give additional evidences
for the existence of the surface-driven electronic structure.

\end{abstract}


\maketitle

\section{Introduction}

The discovery of superconductivity in the iron arsenic
$R$FeAsO$_{1-x}$F$_y$ family \cite{Original} (the ``1111" family,
$R$ being the rare earth elements) has triggered enormous scientific
activity within the last two years. The transition temperature $T_c$
in these materials well exceeds the theoretical maximum predicted by
the Bardeen-Cooper-Schrieffer theory. Though several other families
of iron pnictide superconductors (e.g. the carrier-doped
$A$Fe$_2$As$_2$ or the ``122" family, $A$ being Ba, Sr, Ca, Rh) were
discovered after the initial work,\cite{Original_122, Original_111,
Original_11} the 1111 family still holds the record for the highest
$T_c$ of 55 K.\cite{Ren_55K} There has been an ongoing debate as to
the origin of superconductivity in the iron pnictides as well as
their relation to the traditional copper oxide high-$T_c$
superconductors. Recently Yan \textit{et al.} reported the
successful growth of millimeter-sized LaFeAsO single crystals at
ambient pressure.\cite{Yan_1111} This technical breakthrough brings
experimental studies of these fascinating materials to a new height.

Although angle resolved photoemission spectroscopy (ARPES) has
proven to be a useful experimental method in the field of the
pnictides,\cite{Shen_LaOFeP, Shen_PhysicaC, Liu_1111FS, Ding_gap,
Zhou_gap, Feng_splitting, Terashima_Cogap, Hasan_gap} one major
obstacle has hindered an extensive survey of the 1111 systems.
Apparently the Fermi surface (FS) measured by ARPES is inconsistent
with the theoretical predictions and other experimental probes of
the bulk electronic structure. More specifically, theoretical
calculations and quantum oscillation measurements
\cite{McDonald_dHvA} suggest similar sizes of the $\Gamma$ hole
pockets and the $M$ electron pockets at the FS, while ARPES shows an
extra large circular hole pocket around the zone center $\Gamma$,
which covers almost 40\% of the Brillouin zone intersection
size.\cite{Liu_1111FS, Shen_PhysicaC} Base on the fact that ARPES
can probe only the first layers of the crystal, many authors believe
that the extra hole pocket comes from surface-driven electronic
structure, i.e. an atomic reorganization and/or lattice relaxation
at the sample surface.\cite{Shen_PhysicaC, Mazin_PhysicaC, Eschrig}
The answer to this question is a prerequisite for any further ARPES
investigation of the 1111 system.

The first propose for this paper is to verify the observation of a
surface-driven electronic structure in the 1111 parent compound
LaFeAsO. By comparing with an electronic structure calculation for
the surface layers, we show that indeed the extra $\Gamma$ hole
pocket comes from a surface FeAs or LaO layer. Despite this, the
observation does not exclude the possibility that electronic
structure from the bulk crystal is also present in the \textit{same}
ARPES map. Furthermore, it is likely that such a surface-driven hole
pocket may store the superconducting information of the bulk crystal
via the proximity effect, resulting in e.g. the observation of an
$s$-wave-like superconducting gap by Kondo \textit{et
al}.\cite{Takeshi_gap} Keeping this in mind, we discuss the strong
polarization dependence of the band structure and an unusual
temperature-dependent hole-like band around the $M$ point.

\begin{figure}
\includegraphics[width=3.5in]{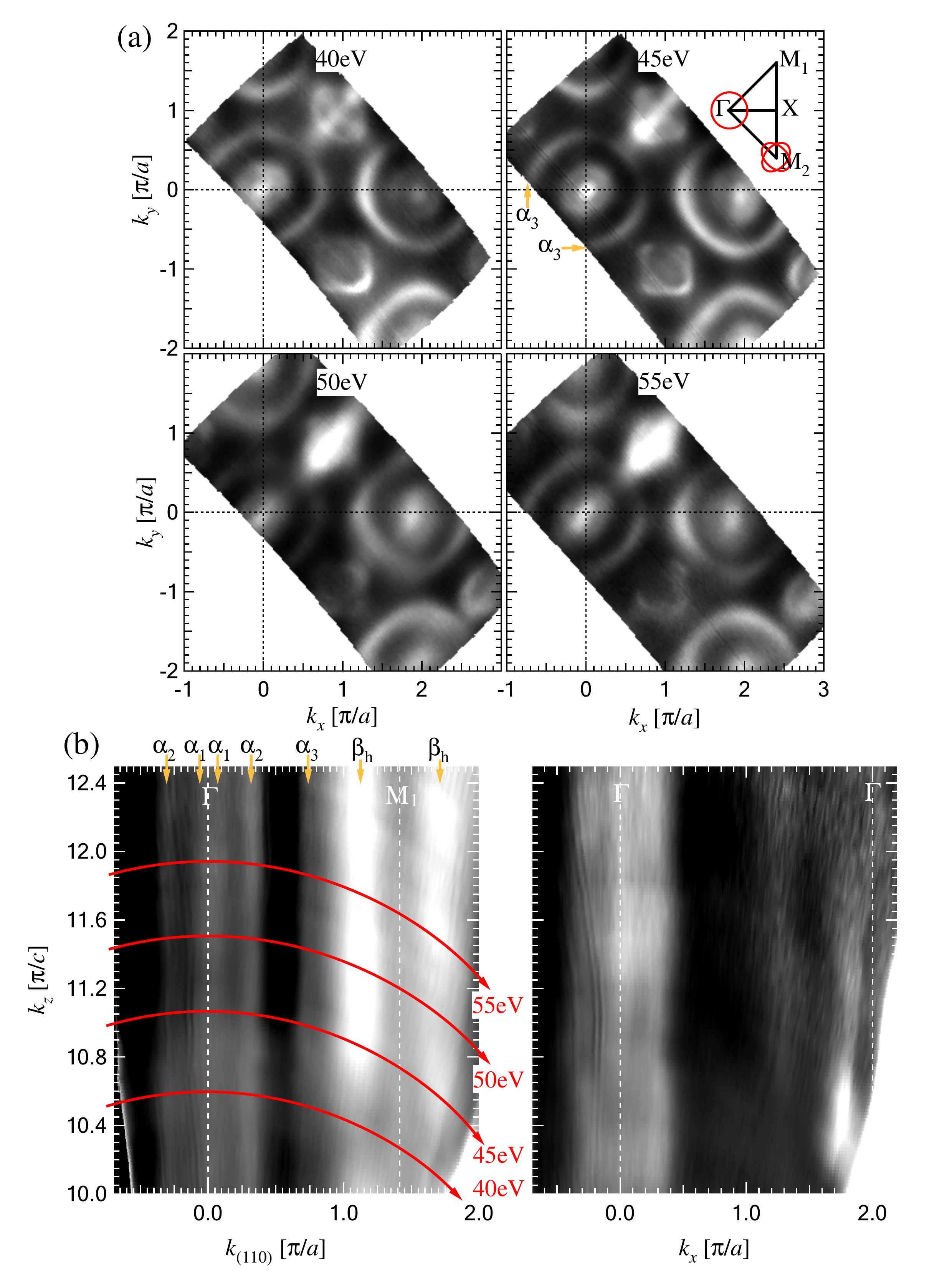}
\caption{(color online) (a) ARPES Fermi surface maps of LaFeAsO,
integrated within $\pm$ 10 meV with respect to the chemical
potential. Bright areas indicate high intensity. Data is taken with
a polarized monochromatic synchrotron beam at $T=10$ K. Incident
photon energies are indicated at the top of each panel. Inset of the
45 eV panel shows the labeling of the high symmetry points. Note
that $M_1$ and $M_2$ represent two zone corners showing different
band structures due to different polarization arrangements. (b)
$k_z$ dispersion maps of LaFeAsO taken with incident photon energies
$30<h\nu<70$ eV for the $\Gamma$-$M_1$ (left) and $\Gamma$-$\Gamma$
(right) directions respectively. The inner potential is 20 eV.
Locations of each Fermi crossing band are indicated by their symbols
and orange arrows. Red arrows show the locations of the four maps in
Fig. 1(a).} \label{fig1}
\end{figure}

\begin{figure}
\includegraphics[width=3.4in]{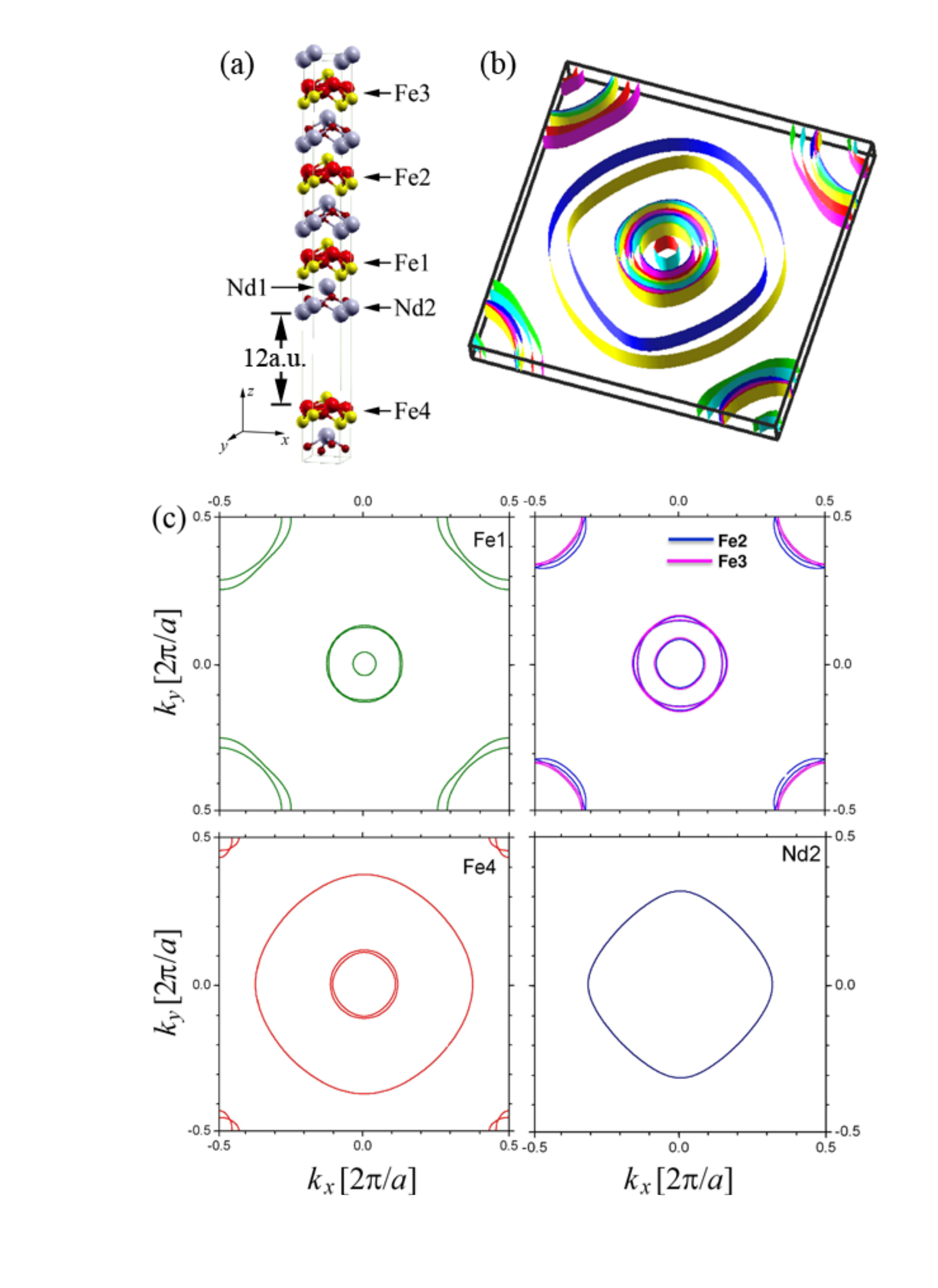}
\caption{(color online) Results of the full-potential Linear
Augmented Plane Wave calculation for the surface-driven electronic
state of NdFeAsO. The results of LaFeAsO will be essentially the
same (see discussion in the text). (a) Calculational setup of the
crystal lattice. The red, yellow, lilac and purple spheres represent
the Fe, As, Nd and O atoms, respectively. Labels such as Fe1 and Nd2
indicate atoms at different locations. The 12 a.u. distance between
Nd2 and Fe4 layers is added manually to imitate the existence of the
crystal surface. (b) Calculation result of the three dimensional
Fermi surface for the crystal in (a). (c) Fermi surface sheets
generated from different atomic layers.
%(d) Calculated band
%structure for the crystal lattice in (a). The bands generated by
%$d$-electrons of the iron atoms located at the Fe4 sites are
%indicated by blue circles. The location of the $\alpha_3$ band is
%specifically pointed out.
} \label{fig2}
\end{figure}

\section{Experimental}

Millimeter-sized single crystals of LaFeAsO were grown out of a NaAs
flux using conventional high-temperature solution growth
techniques.\cite{Yan_1111} As-grown crystals have typical dimensions
of $3\times4\times0.05$ $\sim$ $0.3$ $\textrm{mm}^3$ with the
crystallographic $c$-axis perpendicular to the plane of the
plate-like single crystals. The characteristic temperatures for the
separated structural and magnetic transitions are
$T_\textrm{S}\sim154$ K and $T$* $\sim140$ K, respectively. The
ARPES measurements were performed at beamline 10.0.1 of the Advanced
Light Source (ALS), Berkeley, California using a Scienta R4000
electron analyzer, as well as a laboratory-based ARPES system
consisting of a Scienta SES2002 electron analyzer,
GammaData$^\circledR$ UV lamp and custom designed refocusing optics
at Ames Laboratory. Vacuum conditions were better than
$3\times10^{-11}$ torr. The energy resolution was set at $\sim$ 15
meV for measurements at the ALS, and $\sim$ 9 meV for measurements
at Ames Laboratory. All samples were cleaved \textit{in situ}
yielding mirror-like, clean $a$-$b$ surfaces. Cleaved surfaces of
all samples are stable for at least 24 hours for a given
temperature. High symmetry points $M_1$ and $M_2$ are defined to be
($\pi/a$, $\pi/a(b)$, 0) and ($\pi/a$, $-\pi/a(b)$, 0),
respectively,\cite{symmetry_points} with the $k_x$ and $k_y$ axes
along the Fe-As bonds.

\section{Results and discussion}

Fig. 1(a) shows the ARPES FS maps of LaFeAsO taken with a linearly
polarized synchrotron beam for four different incident photon
energies (probing at four different $k_z$ momenta\cite{Hufner}).
More than half of the first Brillouin zone is covered. The data in
Fig. 1(a) is consistent with previous ARPES studies on LaFePO,
LaFeAsO, CeFeAsO and NdFeAsO$_{0.9}$F$_{0.1}$ single
crystals.\cite{Liu_1111FS, Shen_PhysicaC, Zhou_CeFeAsO} The most
apparent difference between the ARPES data on these 1111 systems and
that on the 122 systems is a ultra large Fermi pocket around
$\Gamma$, which is denoted as $\alpha_3$ throughout the paper.
Subsequent band structure analysis [Fig. 3(a)] shows that this
pocket is hole-like. This pocket is not expected from theoretical
calculations\cite{Liu_1111FS} and experimental results for the bulk
electronic structure.\cite{McDonald_dHvA} Recall that the electron
escape depth for ARPES experiments are of the order of only a few
angstroms (comparable with the lattice constant $c$), ARPES is
essentially a surface probe.\cite{Hufner} It is thus reasonable to
speculate that the existence of the $\alpha_3$ pocket is a result of
a surface-driven electronic structure. The $k_z$ dispersion maps in
Fig. 1(b) give further evidence for this speculation. It can be
easily seen from Fig. 1(b) that the detected electronic structure,
including the $\alpha_3$ pocket, is essentially two dimensional
along both $\Gamma$-$\Gamma$ and $\Gamma$-$M$ directions. This
observation is in sharp contrast with the three-dimensional
electronic structure observed in 122 compounds,\cite{Liu_3D, Brouet}
while consistent with the two dimensional nature of the sample
surface layer.

In Fig. 2 we present the results of a model theoretical calculation
to verify the existence of the surface-driven electronic structure.
It should be noted here that although this calculation is performed
for NdFeAsO, the results for LaFeAsO will be essentially the same
(see also Ref. \cite{Eschrig}), since the 4$f$ electrons of the Nd
atoms were treated as core electrons, the valence electrons of the
Nd atoms are the same as those of the La atoms. In this calculation
we use a full-potential Linear Augmented Plane Wave (FPLAPW)
method\cite{LAPW} with a local density functional.\cite{LDA} The
structural data was taken from a reported experimental
result.\cite{Qiu} The presence of the crystal surface is imitated by
constructing a supercell with four NdO and FeAs layers, and a 12
a.u. vacuum located between the Nd2 and Fe4 layer [Fig. 2(a)]. To
obtain the self-consistent charge density, we chose 28 $k$-points in
the irreducible Brillouin zone, and set $R_{\textrm{MT}}\times
k_{\textrm{max}}$ to 7.5, where $R_{\textrm{MT}}$ is the smallest
muffin-tin radius and $k_{\textrm{max}}$ is the plane-wave cutoff.
We use the muffin-tin radii of 2.4, 2.1, 2.1 and 1.6 a.u. for Nd,
Fe, As and O respectively. In this calculation, the atoms near the
surface (Nd1-O-Nd2, As-Fe4-As) were relaxed along the $z$-direction
until the forces exerted on the atoms were less than 2.0 mRy/a.u.
With this optimized structure, we obtained a self consistency with
0.01 mRy/cell total energy convergence. After that, the three
dimensional Fermi surface calculation was performed with 420
$k$-points in the irreducible Brillouin zone [Fig.
2(b)].\cite{XcrysDen} For the two dimensional electronic structure,
we chose $k_z=0.5$ (the $k_x$-$k_y$ plane that crosses the $\Gamma$
point) and divided the Brillouin zone ($-0.5 < k_x$, $k_y < 0.5$)
into a $51\times51$ mesh that resulted in 2601 $k$-points [Figs.
2(c) and 2(d)].

From Figs. 2(b)-(d), one essential statement must be pointed out:
The surface layer of NdFeAsO (and LaFeAsO), regardless of its
elemental nature (NdO/LaO or FeAs), will generate an extra large
hole pocket at the Fermi surface, and this large hole pocket can
only be generated at the surface layer of the Nd/LaFeAsO crystal. In
contrast, even though it is just the second layer from the surface,
the Fe1 layer generates two $\Gamma$ pockets which are even smaller
than the $M$ pockets from the same layer. Similar sizes of the
$\Gamma$ and $M$ pockets are seen only from the Fe2 and Fe3 layers,
which are farther away from the surface and can thus be considered
``bulk" states. Assuming the validity of the rigid band shifting
scheme, this effect can be explained by the transfer of charge near
the surface: electrons are retrieving from the surface layer (Fe4 or
Nd2) to the adjacent layer (Fe1), leaving the opposite sign of the
electron-hole imbalance for these two layers.

\begin{figure}
\includegraphics[width=3.2in]{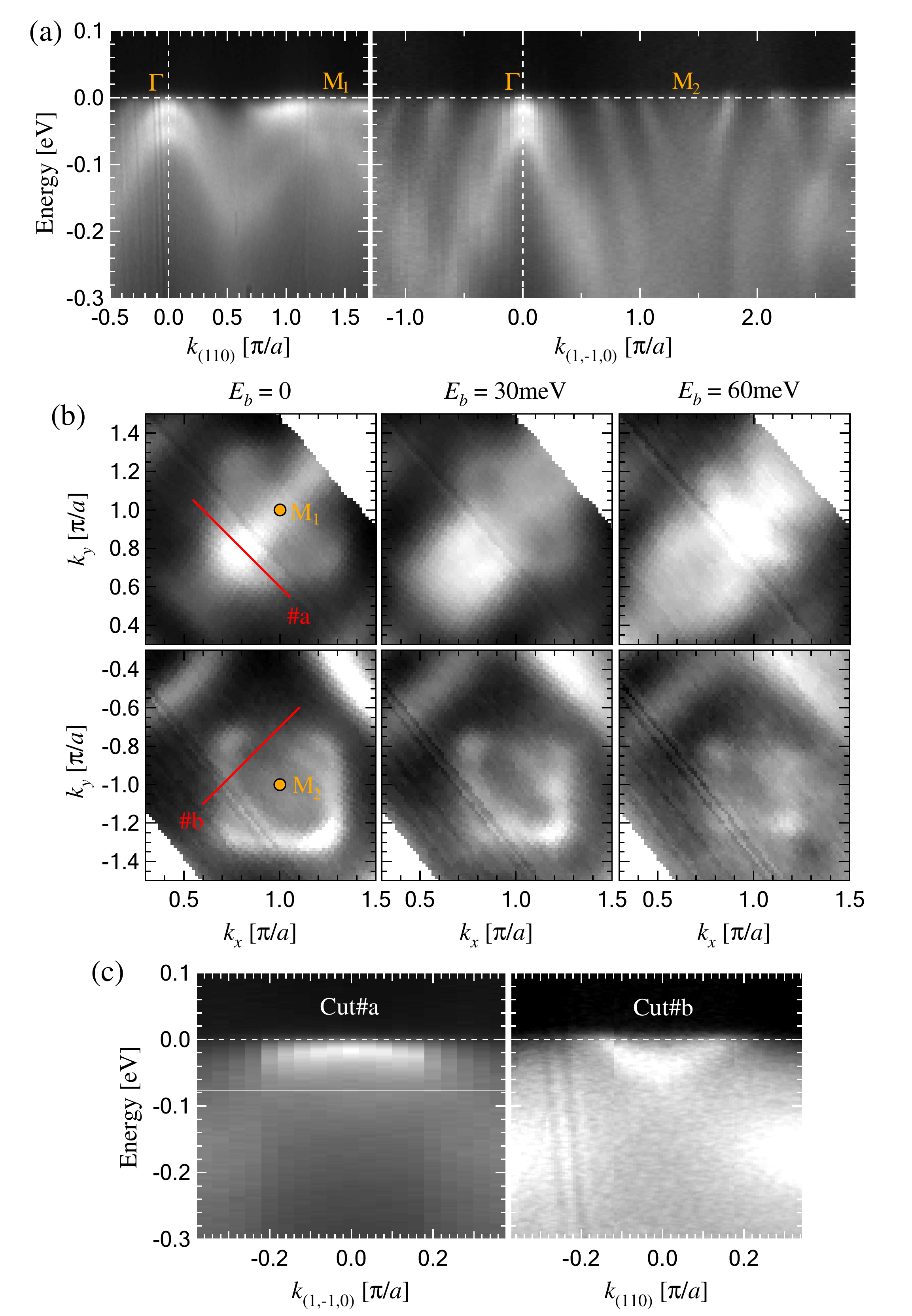}
\caption{(color online) Band structure analysis for the 45 eV data
in Fig. 1. (a) Measured band structure for two diagonal cuts
$\Gamma$-$M_1$ and $\Gamma$-$M_2$. Strong polarization dependence on
the electronic structure in both the $\Gamma$ and $M$ pockets are
clearly visible. (b) Constant energy maps for binding energies
$E_b=0$, 30 and 60 meV in the vicinity of $M_1$ and $M_2$ points.
Binding energies are indicated to the top of each column. Cut\#1 and
\#2 indicate the locations of the band structure maps in panel (c).
} \label{fig3}
\end{figure}

\begin{figure}
\includegraphics[width=3.3in]{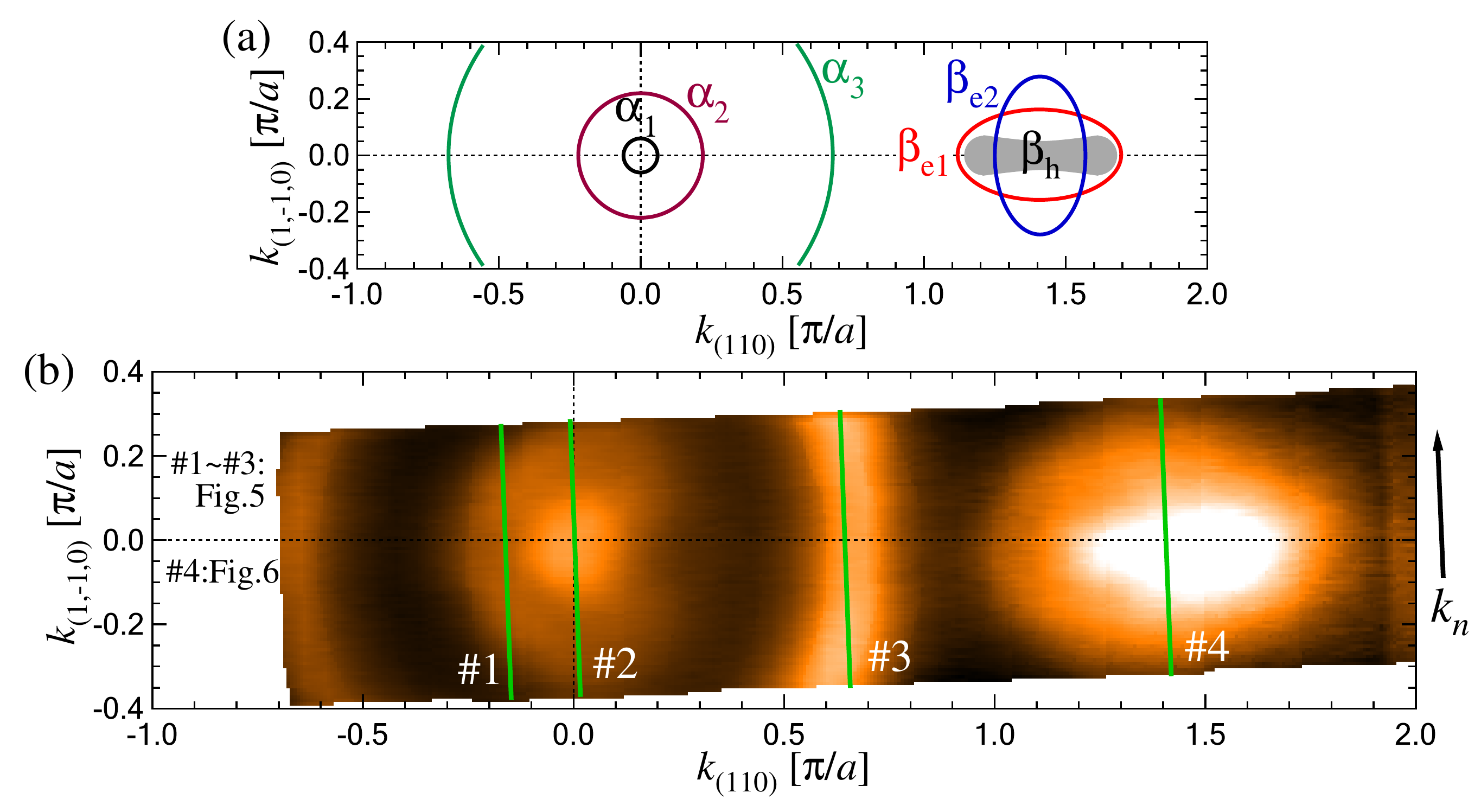}
\caption{(color online) (a) Notations of the Fermi pockets and bands
of LaFeAsO measured by ARPES. The existence of these pockets/bands
is proved by the data from the synchrotron (Fig. 1 and 3). The three
hole pockets around $\Gamma$ are labeled as $\alpha_1$, $\alpha_2$
and $\alpha_3$; $\beta_{\textrm{e1}}$ ($\beta_{\textrm{e2}}$) is the
$M$ electron pocket with the long axis along (perpendicular to) the
$\Gamma$-$M$ direction; $\beta_{\textrm{h}}$ represents the
intensity from a hole-like band located right below the chemical
potential $\mu$ at the vicinity of $M$. (b) ARPES intensity map at
$\mu$ measured with a helium lamp ($h\nu=21.2$ eV), along with
locations of the ARPES band structure cuts shown in Figs. 5 - 7. The
direction of these cuts is defined as $k_n$.} \label{fig4}
\end{figure}

Another interesting observation of Fig. 1 is the strong polarization
dependence of the $M$ electron pockets. As seen from Fig. 1(a), the
two $M$-points $M_1$ and $M_2$ which are $90^\circ$ away in the
$k$-space show fundamentally different electronic structures. Fig. 3
shows their detailed band structure analysis. Clearly the $M_1$
point is surrounded by four petal-like intensity peaks at the
chemical potential. The ones along the $k_{(110)}$ direction [Cut\#a
in Fig. 3(c)] are much more pronounced in intensity, and they are
indeed hole-like bands, the top of which located at $\sim$20meV
below the chemical potential. We denote this band as
$\beta_\textrm{h}$ [Fig. 4(a)]. The intensity peaks along the
$k_{(1,-1,0)}$ direction are actually part of an elliptical electron
pocket. On the other hand, the bands around the $M_2$ point manifest
themselves as two crossed elliptical electron pockets [Cut\#b in
Fig. 3(c)]. We denote them as $\beta_{\textrm{e1}}$ and
$\beta_{\textrm{e2}}$ [Fig. 4(a)]. We speculate that one of these
electron pockets - the one whose long axis is perpendicular to the
$\Gamma$-$M_2$ direction, $\beta_{\textrm{e2}}$ - is the same pocket
as the one observed around $M_1$.
%This unusual behavior is neither
%due to the antiferromagnetic order nor the orthorhombic nature of
%the crystal, since (1) if we rotate the polarization direction of
%the incident beam by $45^\circ$, both $M_1$ and $M_2$ show a
%combination of the small hole bands and the larger crossed electron
%pockets (not shown); and (2) if we measure the same sample under a
%partially polarized helium lamp beam, we can observe intensities
%from all three bands/pockets [see Fig 4(b) and the following
%discussion].
The only way to explain this unusual behavior is that these
$M$-pockets are highly sensitive to the polarization of the incoming
beam. In fact such phenomena are observed and discussed in detail in
the 122 systems.\cite{Hasan_pol, Feng_pol} In Fig. 1 and Fig. 3 in
this paper, the electric field vector of the incoming light is
polarized along the $k_{(110)}$ direction. The entrance slit of the
electron analyzer is perpendicular to the mirror plane defined by
the incident beam and the sample surface normal.
%Following the
%notation by Hsieh \textit{et al.},\cite{Hasan_pol} what we find in
%this 1111 parent compound is that the $\beta_\textrm{h}$ band is
%only visible under the $\textbf{E}_{\textit{s}}$ geometry, it has
%parity-odd orbital symmetry ($d_{yz}$ or $d_{xy}$). The
%$\beta_{\textrm{e1}}$ pocket is only visible under the
%$\textbf{E}_{\textit{p}}$ geometry, it has parity-even orbital
%symmetry ($d_{xz}$, $d_{x^2-y^2}$ or $d_{z^2}$). The
%$\beta_{\textrm{e2}}$ pocket is visible under both
%$\textbf{E}_{\textit{s}}$ and $\textbf{E}_{\textit{p}}$ geometry, so
%it has both parity-odd and even components.
The fact that the
$\beta_{\textrm{e1}}$ and $\beta_{\textrm{e2}}$ pockets have
different parity nature is consistent with tight-binding
calculations.\cite{Graser}

%Now we summarize the results of the ARPES measurements using a
%polarized synchrotron beam. We observe three hole pockets centered
%at the zone center $\Gamma$, denoted as $\alpha_1$, $\alpha_2$ and
%$\alpha_3$ with increasing sizes [see Fig. 4(a)]. By comparing with
%theoretical calculation we conclude that at least the largest hole
%pocket $\alpha_3$ is a result of a surface-driven electronic
%structure.

\begin{figure}
\includegraphics[width=3.5in]{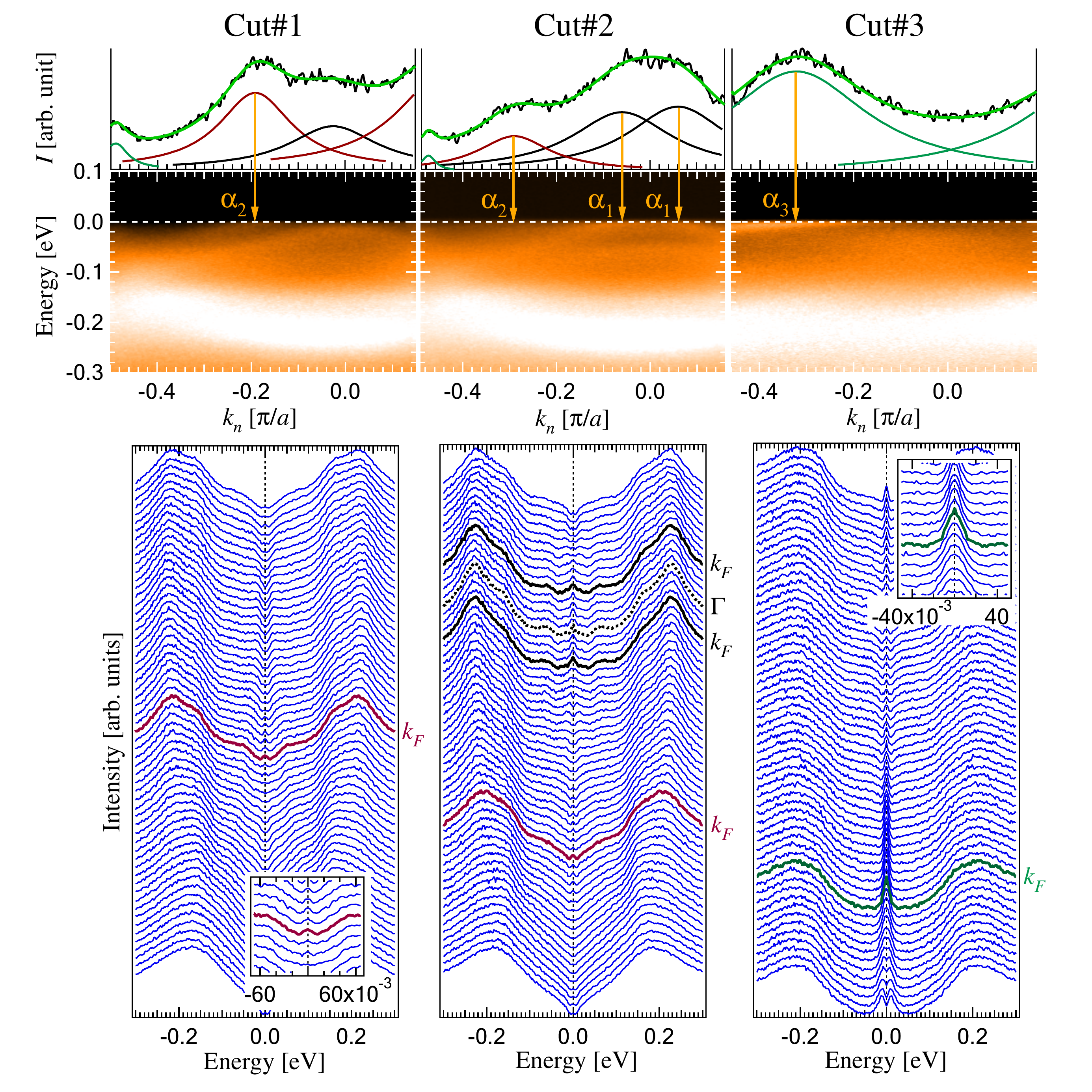}
\caption{(color online) Detailed band structure analysis for the
$\Gamma$ hole pockets. Locations of the cuts \#1, \#2 and \#3 are
indicated in Fig. 4(b). Top row: Momentum distribution curves (MDCs)
integrated within $\pm$ 10 meV with respect to the chemical
potential $\mu$ for the three cuts. $\mu$ is determined by fitting a
Fermi function to the spectra of polycrystalline gold. The MDCs are
fitted with several Lorenzians for extracting the peak positions
which are used to determine the Fermi crossing momenta ($k_F$s).
Middle row: Band dispersion maps obtained by ARPES measurements at
$T=12$ K. The incident photon energy is 21.2 eV. Bottom row: Energy
distribution curves (EDCs) for the corresponding maps in the middle
row. Each EDC is integrated within a momentum range of 0.011$\pi/a$
and is symmetrized with respect to $\mu$. EDCs corresponding to the
Fermi cross momenta of each pocket are marked by the same color used
for Fig. 4(a). Insets show an expended region at the vicinity of the
$k_F$s.} \label{fig5}
\end{figure}

We now move on to a detailed discussion of how these bands/pockets
behave at the Fermi level. The high energy resolution data is taken
with a partially polarized helium lamp ($h\nu=21.2$ eV). We begin
this discussion with Fig. 5 where we analyze three $k$-$E$ maps
marked \#1, \#2 and \#3 in Fig. 4(b). These maps are chosen so that
all three $\Gamma$ hole pockets $\alpha_1$, $\alpha_2$ and
$\alpha_3$ are visible. The main conclusion of this figure is that
there is no gap opening at the $\Gamma$ pockets. In the symmetrized
energy distribution curves (EDCs) at each Fermi crossing momenta, we
see one single peak instead of two peaks separated by a valley. This
is a typical indication of a vanishing gap. Linking this result with
the findings by Kondo \etal,\cite{Takeshi_gap} we obtain an
interesting picture for the low lying electronic excitations of this
1111 system. Resulted solely from the surface layer, the $\alpha_3$
pocket shows no gap in the undoped sample, while exhibits an
$s$-wave-like gap as large as $\sim15$ meV in the fluorine-doped
superconducting sample.\cite{Takeshi_gap} The coherent peak for this
gap vanishes right at $T_c$.\cite{Liu_1111FS} To date we have no
information on whether the surface layer of the 1111 crystal is
superconducting itself, so a possible scenario is the proximity
effect. The proximity effect offers us the possibility that the
superconducting properties of the bulk sample affects the surface.
In that case the momentum dependence of the superconducting gap
observed for the surface will be similar to that of the bulk
crystal. For high-$T_c$ cuprates, it is proposed that the proximity
effect may result in ``$d$+$s$"-wave superconductivity at the
metallic layer that is coated on the superconductor.\cite{Kohen,
Sharoni, Lofwander} This scenario is consistent with scanning
tunneling spectroscopy (STS) results on polycrystalline 1111 samples
where a reduced superconducting gap is observed at the sample
surface.\cite{Millo, Pan} It is not understood why the gap size
shown by ARPES is much bigger than by STS and other methods.

%\begin{figure}
%\includegraphics[width=2.8in]{Fig6.eps}
%\caption{(color online) Band structure analysis for the $M$ electron
%pocket $\beta_{\textrm{e1}}$ at low and high temperatures. Location
%of the cut \#4 is indicated in Fig. 4(b). This figure is arranged in
%the same way as Fig. 5. Different symbols in the EDCs of the low
%temperature data represent different visible bands. Note that the
%electron pocket $\beta_{\textrm{e1}}$ is marked by solid circles and
%the hole-like band $\beta_\textrm{h}$ is marked by hollow
%triangles.} \label{fig6}
%\end{figure}

\begin{figure}
\includegraphics[width=3.3in]{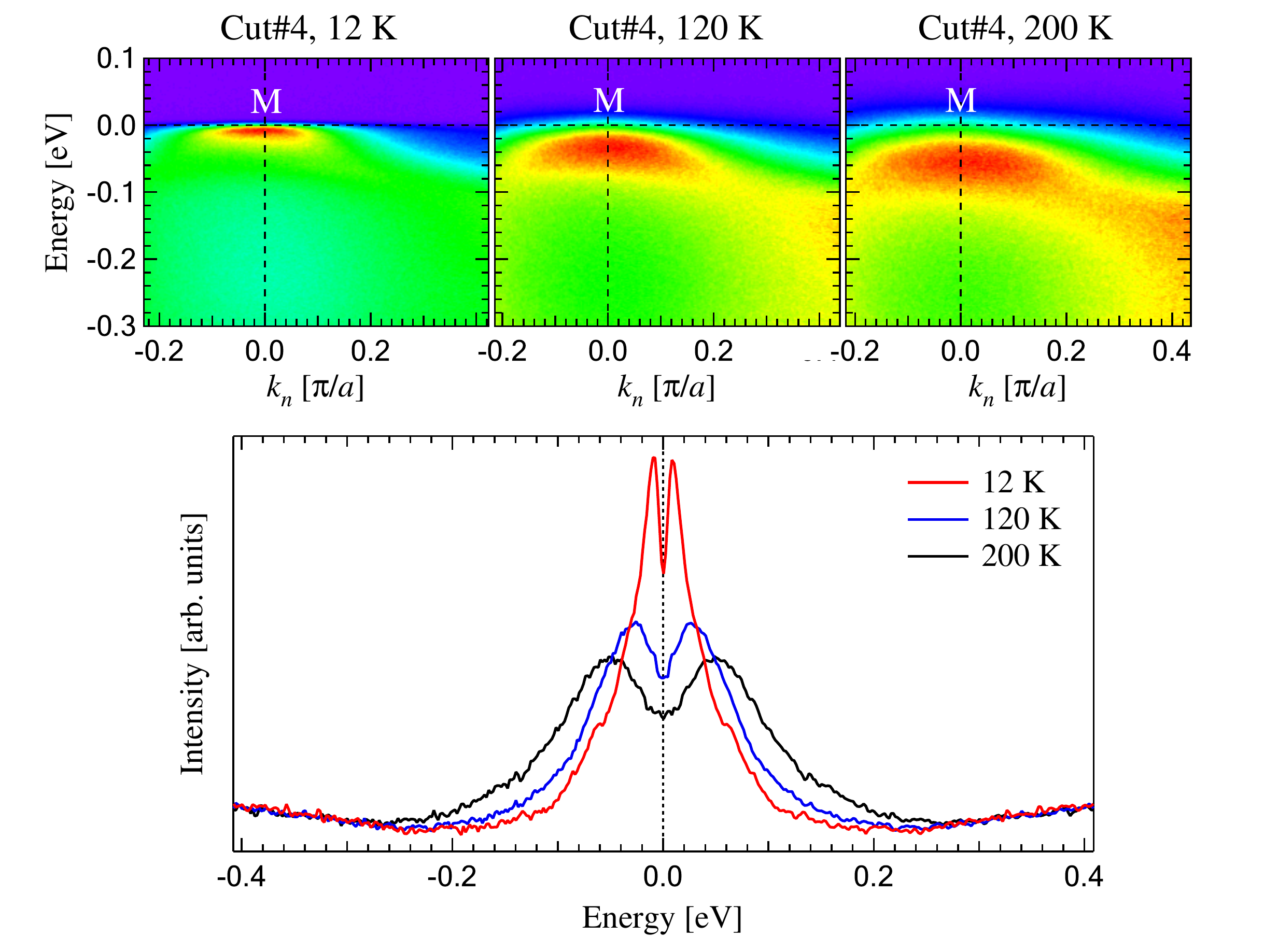}
\caption{(color online) Band structure analysis for the $M$ hole
band $\beta_{\textrm{h}}$ at three different temperatures. Location
of the cut \#4 is indicated in Fig. 4(b). Top row: Band dispersion
maps obtained by ARPES measurements at $T=12$, 120 and 200 K. The
incident photon energy is 21.2 eV. Vertical dashed lines indicate
the location of the high symmetry point $M$. Bottom row: EDCs at the
$M$ point for the three temperatures. Data is integrated within a
momentum range of 0.011$\pi/a$, symmetrized with respect to $\mu$,
and normalized for high binding energies.} \label{fig7}
\end{figure}

In Fig. 6 we discuss the temperature dependence of the
$\beta_{\textrm{h}}$ band. Cut\#4 is located right through the $M$
point where the binding energy is the lowest for this band. One can
easily see from Fig. 6 that the peak location of the
$\beta_{\textrm{h}}$ band gradually shifts to higher binding
energies with increasing temperature. At $T=12$ K the maximum
intensity is located at $E_b\sim10$ meV, whereas at $T=200$ K it
shifts to $E_b\sim60$ meV. The peak intensity decreases at the same
time. Such changes are intrinsic, since multiple measurements are
performed in multiple samples, and both increasing and decreasing
the temperature during the measurement reveals the same behavior. As
a reminder, in the 122 systems the bands locate at the same binding
energies until the magnetic transition temperature is
reached.\cite{Feng_splitting, Liu_3D, Zhou_recon, Shen_recon} Here
the band location shifts even below the transition temperature
(between 12 and 120 K). This is not expected from the bulk
properties - transport measurements unambiguously show a magnetic
transition at 140 K. It is thus another evidence for the existence
of a temperature-dependent surface layer.

\section{Conclusions}

The main conclusion of this paper is that the surface-driven
electronic structure plays an important role in the ARPES data of
the iron arsenic 1111 systems. It is a fact that ARPES data on these
systems shows more discrepancies rather than agreements with
theoretical and experimental \textit{bulk} properties. ARPES shows a
temperature-dependent $\beta_\textrm{h}$ band, and most
significantly an extra large hole pocket around $\Gamma$. However,
bulk calculations and experiments suggest a Fermi surface
reconstruction associating with a well-defined magnetic transition
temperature, and similar sizes of $\Gamma$ and $M$ pockets. By
comparison with a full-potential Linear Augmented Plane Wave
calculation on the surface layers of the crystal, we confirm that at
the very least the large $\Gamma$ pocket is generated solely by the
sample surface. The $s$-wave-like superconducting gap that exists in
this pocket in the superconducting samples is most likely a result
of the superconducting proximity effect.

\section{Acknowledgments}

We thank S.-K. Mo for his grateful instrumental support at ALS. Ames
Laboratory was supported by the Department of Energy - Basic Energy
Sciences under Contract No. DE-AC02-07CH11358. ALS is operated by
the US DOE under Contract No. DE-AC03-76SF00098.

\end{document}